\documentclass[lettersize,journal]{IEEEtran}
\usepackage{amsmath,amsfonts}
\usepackage{algorithmic}
\usepackage{array}
\usepackage[caption=false,font=normalsize,labelfont=sf,textfont=sf]{subfig}
\usepackage{textcomp}
\usepackage{stfloats}
\usepackage{url}
\usepackage{verbatim}
\usepackage{graphicx}
\usepackage{cite}
\usepackage{bm}
\usepackage{mathrsfs}
\usepackage{color}

\usepackage[ruled,vlined]{algorithm2e}

\begin{document}

\title{Aiming in Harsh Environments: A New Framework for Flexible and Adaptive Resource Management}

\author{Jiaqi Zou, Rui Liu, Chenwei Wang, Yuanhao Cui, Zixuan Zou, Songlin Sun, Koichi Adachi
\thanks{Jiaqi Zou, Yuanhao Cui, Zixuan Zou, and Songlin Sun (corresponding author) are with Beijing University of Posts and Telecommunications, Beijing, China. 

Rui Liu is with San Jose State University, San Jose, California, US. 

Chenwei Wang is with Google, LLC, Mountain View, California, US. This work was performed when he was with DOCOMO Innovations, Inc., Palo Alto, California, US. 

Koichi Adachi is with the University of Electro-Communications, Tokyo, Japan.}
}



\maketitle

\begin{abstract}
The harsh environment imposes a unique set of challenges on networking strategies. In such circumstances, the environmental impact on network resources and long-time unattended maintenance has not been well investigated yet.
To address these challenges, we propose a flexible and adaptive resource management framework that incorporates the environment-awareness functionality. In particular, we propose a new network architecture and introduce the new functionalities against the traditional network components. The novelties of the proposed architecture include a deep-learning-based environment resource prediction module and a self-organized service management module. Specifically, the available network resource under various environmental conditions is predicted by using the prediction module. Then based on the prediction, an environment-oriented resource allocation method is developed to optimize the system utility. To demonstrate the effectiveness and efficiency of the proposed new functionalities, we examine the method via an experiment in a case study. Finally, we introduce several promising directions of resource management in harsh environments that can be extended from this paper.

\end{abstract}


\section{Introduction}
\IEEEPARstart{H}{arsh} environments for humans are considered as environments with extreme conditions, which are difficult or impossible for humans to survive. This description can also be extended to electronic communication devices, which in the harsh environment would experience significant degradation of network performance. As a result, even meeting the basic communication demands without special designs might be questionable. 
The harsh environment is usually unattended and dominated by multiple severe environmental factors, as well as the network itself. For example, in the iron and steel manufactory, the network is expected to tolerate with extreme temperatures, vibration, high pressure, moisture and noxious gases, etc; in the extreme natural environment, 
the network should take into account appalling weather, such as extreme drought, heavy rain and mechanical damage.

Although specially-designed electronics that are made by high temperature resistance and non-corrosion materials could survive under extreme conditions \cite{mihailov2012fiber}, the resource management strategies in the harsh environment are much more complicated.
In a future wireless network, resource management is recognized to play an essential role due to the high-demand but limited network resources \cite{ olwal2016survey}. Recent development in \cite{tang2020Deep} provides an adaptive network resource allocation algorithm in high-mobility communication systems. Framework designs based on software defined network also enable scalable resource management, such as the work on space-air-ground integrated communications \cite{wu2020resource}. While these works successfully addressed the issues in a few scenarios, they still lack flexibility under changeable environments. In particular, resource management under extreme physical conditions has not been fully investigated. Also, in the harsh environment, the traditional resource management systems would encounter many new challenges, because the dynamic harsh conditions could significantly affect network performances, and of the incapability of autonomously recovering reliable communication services under long-time unattended conditions.


When network resource management meets the harsh environment, there are two main challenges in general:
\begin{enumerate}
    \item {\it Long-term reliable, autonomous service management with quick responses under the unattended environment:} In-person maintenance is usually expensive or even impracticable for the network in the harsh environment. Thus, self-healing and self-maintenance capabilities are expected to be integrated, so that the services and resources could be autonomously formed and managed to guarantee the reliability and stability. For instance, the traditional network architectures, e.g., ad hoc \cite{ramanathan2002brief}, provide a flexible topology, but they lack reliability in unattended and extreme environments. When the main link fails, the ongoing transmission would be re-routed to sub-optimal backup links, which would temporarily depress the network performance. In the harsh environment, the network actually requires flexible and adaptive configuration of network resources. Moreover, the scalability of the service management is also a key to build fast and durable solutions.
    
    \item {\it Consideration of the uncertainty impact on network resources caused by complicated environmental factors:} Compared to the normal environments where the environmental limitations have certain distribution, such as path loss and rain attenuation, the status of network resources in the harsh environment is affected and dominated by multiple environmental factors. These factors are commonly coupled and environment-dependent, so that mathematical modeling is intricate. 
    Moreover, harsh conditions often seriously suppress the network performance. Thus, the environmental impact needs to be specially considered when designing a network resource management framework.
\end{enumerate}

Recognizing the challenges above, in this paper, we propose a flexible and adaptive resource management framework that consists of a new resource management architecture and an environment-oriented resource allocation method. Specifically, to achieve autonomous service management, the proposed architecture realizes self-organization and self-maintenance of the services; to combat complicated environmental factors, an deep-learning-based model to predict the environmental impact on available network resources is integrated into the proposed resource management architecture, and the prediction of available network resources is taken into account in the proposed resource allocation method. 

The remainder of this paper is organized as follows. In Section \ref{sec:architecture} we propose an environment-aware self-maintenance resource management architecture and introduce its reliability, scalability and the environment perception module. In Section \ref{sec:framework}, we formulate an optimization problem considering resource allocation under harsh conditions. To resolve it, in Section \ref{sec:solution}, we first design a deep-learning-based method for available resource prediction and then develop a game-theory-based method for resource allocation. To illustrate the effectiveness of the proposed architecture, we provide a case study in Section \ref{sec:example}. Finally, we conclude this paper with the potential future research directions in Sections VI and VII.

\section{Resource Management: From the Traditional to the New Architecture}

\begin{figure*}[!t]
\centering
\includegraphics[width=0.95\textwidth]{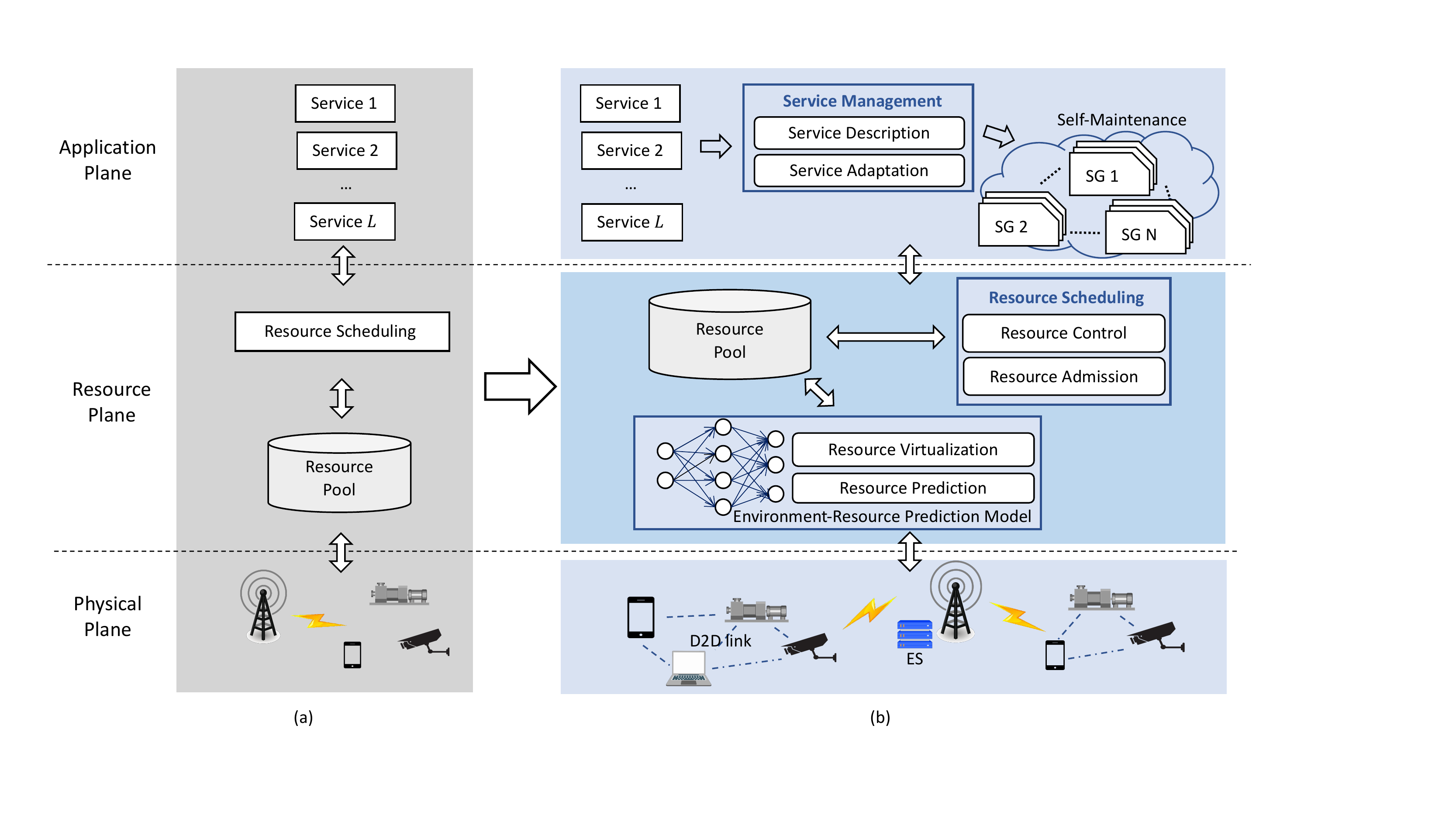}\vspace{-0.1in}
\caption{(a) The traditional resource management architecture. (b) Our proposed environment-aware self-maintenance resource management architecture.}
\label{arc}
\end{figure*}

The resource allocation in the wireless network is implemented in the resource management architecture, which has been widely investigated in literature. Although each architecture has its own attributes, a structure of three planes -- application, resource, and physical planes -- are usually formed, as shown in Fig.~\ref{arc} (a). In the following, we briefly introduce the function of each plane.
\begin{itemize}
    \item \emph{The physical plane} aggregates and controls the physical devices distributed over the space. For collaborative sensing of environment conditions and collecting real-time information, a vast amount of the distributed sensors should be deployed.
    \item \emph{The resource plane} processes the information sent from the physical plane and allocates network resources based on the carefully-designed criteria of interest. 
    \item \emph{The application plane} supports a variety of services, such as intelligent monitoring, remote controlling and cloud computing. The services can be task-specific applications and public platform services.
\end{itemize}

The existing architectures, e.g., in \cite{zhang2017hierarchical}, decouple the service management logic in the application plane and the control logic in the resource plane. Such hierarchical architectures are widely adopted and enable the network management to operate from a global view of a unified resource plane \cite{nunes2014survey}. Similar architectural paradigms are also considered by forward-looking research in 6G networks with abundant emerging applications \cite{huang2019survey}. However, these architectures of network resource management are usually statically deployed, which are not capable of perceiving environment and responding to the dynamic harsh conditions. Also, how to autonomously maintain reliable services under long-time unattended conditions appears very difficult. Thus, designing a new architecture in the presence of hard conditions would be desired.

\subsection{The Proposed New Architecture}\label{sec:architecture}

We propose a new environment-aware and self-maintenance architecture for resource management, as illustrated in Fig.~\ref{arc} (b). There are two main novelties in the new architecture: (1) \emph{the prediction module} in the resource plane, which is for environment sensing based on a pre-trained data-driven model; (2) \emph{the service management} in the application plane, which is autonomous, self-maintained and scalable to support the adaptation between the service requirements and the available resource amount. In this section, we introduce the function flow of the new architecture in each plane in detail.

Firstly, in the \emph{physical plane}, the harsh environment is an ideal use case for more capable sensors and edge computing. Different from the sensors for the normal environment, the sensors designed for the harsh environment have larger storage capacity to prepare for communication-link failure. Moreover, the edge-computing server (ES) can reduce the communication delay and the bandwidth usage owing to local processing. The devices can also communicate with each other with the help of device-to-device (D2D) links, thus providing emergency communications in case of natural disasters, e.g., an ad hoc network can be established via D2D. To boost the reliability, the devices are distributively deployed to attenuate the effect of individual failure\footnote{The network stability under the harsh environment requires reliable transmission and access protocols. Hence, the alarm report transmission with very high reliability is necessary when an extreme event occurs.}. As a result, the physical plane collects the device information and the environment condition, which are then fed to the resource plane for efficiently making a more accurate decision for controlling network resources. 

Second, in the \emph{resource plane}, the proposed environment-resource prediction module interacts with the physical plane and the resource pool to assure the compliance of the propagated resource schedule request. Specifically, it receives the environment information as the input and responds to resource pool with the resource-allocation decision. The resource pool is a logical component for flexible management of resources. The available resources, such as bandwidth and power budget, are aggregated and shared by multiple services. To learn the underlying patterns behind the environment features and the output system metrics, we leverage deep learning to monitor, predict and adapt to the resource status. Then the available network resources predicted by prediction model are aggregated in the resource pool for resource scheduling. Note that compared to the traditional resource scheduling, two functionalities are highlighted in the new scheduling module -- the resources are authenticated and authorized in the resource admission sub-module, and then the resource control sub-module takes an action of power control, load control, and radio resource control, etc. Meanwhile, the resource scheduling also takes the environment constraints predicted by environment-resource prediction module.


Finally, in the \emph{application plane}, the self-organization and self-maintenance of services are realized. To achieve this goal, we design the service management module which consists of the description and the adaptation sub-modules, as shown in Fig.~\ref{arc} (b). The service-description one describes the resource requirements from the services, and then the service-adaptation one makes an adaptive orchestration to generate several service groups (SGs) by grouping the services with similar functions (but their required types of network resources can be different). Owing to this graph-based structure, whenever some (but not all) services are down due to the resource limitation, other services in the same SG would immediately back up. When the report from the resource plane changes, the SG organization can also be updated and optimized according to the dynamic resource states. In addition, if some services cannot be supported due to the resource changes, new SGs can be re-organized. As a result, self-organization and self-maintenance can be achieved in a certain level. 

\section{The Problem Formulation of Resource Allocation Under Harsh Conditions} \label{sec:framework}

Based on the flexible resource management architecture proposed in Section \ref{sec:architecture}, we design a new framework of environment-oriented resource allocation in response to the dynamics of the harsh environment. The framework leverages the environment-aware resource prediction to maximize the network utility function of interest.

To characterize the problem of resource allocation under the harsh environment, let us consider a network with $D$-dimensional resources supporting a total of $L$ services. Meanwhile, we define the following three notations. 
\begin{itemize}
    \item ${\bf A}$ denotes the resource allocation matrix where the entry $a_{l,d}$ represents the resource quantity allocated to the $l$-th service on the $d$-th resource.
    
    \item ${\bf W}$ is a \emph{predefined} attention matrix with $L$ rows and $D$ columns, where the entry $w_{l,d}$ characterizes the amount of attention that the service $l$ pays to the $d$-th resource, and the sum of each row of ${\bf W}$ equals one for the normalization purpose. 
    
    \item ${\bf r} = [r_1, r_2, \cdots, r_D]$, a $D$-dimensional vector, denotes the amount of the available network resources, such as bandwidth and throughput. Note that in the harsh environment, ${\bf r}$ is \emph{environment-dependent}. 
\end{itemize}

{\it A Toy Example:} For simplicity, consider $L=3$ services and $D=2$ types of the network resource, where 
\begin{eqnarray*}
{\bf A} = \left[\begin{array}{rr}6~\text{Mbps}&25~\text{Mb}\\ 10~\text{Mbps}&40~\text{Mb} \\ 1~\text{Mbps}&100~\text{Mb}\end{array}\right], ~~~
{\bf W} = \left[\begin{array}{rr}0.3&0.7\\ 0.8&0.2\\ 0.4&0.6\end{array}\right].
\end{eqnarray*}
The two columns of ${\bf A}$ have different units, representing the allocated rates and bandwidth in the service layer, and ${\bf W}$ indicates the preference levels, i.e., the weights, of the corresponding allocated resources.  

With the notations described above, we formulate the problem of resource allocation to maximize a sum of utility with network resource constraints. We define the utility function of interest as $w_{l,d}f(a_{l,d})$, where $f(\cdot)$ denotes a mapping function from the allocated resource quantity to the quality-of-service (QoS) requirements\footnote{For instance, according to the Open System Interconnection (OSI) Model, during the initial Transmission Control Protocol (TCP) handshake procedure, the max acceptable establishment delay is inherently associated with the failure probability of the connection establishment. The forms of $f(\cdot)$ can be flexibly defined according to the service types, and some widely used forms include linear function, a step function, and non-linear functions.}. The allocated resource quantity follows $\sum_{l=1}^L a_{l,d} \leq r_d$ that denotes the resource budgets. Note that if $f(\cdot)$ is convex, the problem can be \emph{efficiently} solved by using the standardized inner-point method in polynomial time of the dimensions $L$ and $D$. However, if $f(\cdot)$ is non-convex, which is very common in practice, then the optimality cannot be efficiently achieved or even discovered. Hence, it is necessary to design an efficient method to explore the (sub-)optimality.


In our proposed architecture, incorporating the use of the vector $\bf r$ is a key, which is different from the traditional architectures. It can be inferred by the proposed environmental perception module in response to the harsh environment changes. To do it, we need to investigate how to well predict the available network resources based on the perception of the environmental conditions\footnote{The underlying mapping, if available, is usually non-linear and not straightforward, because a few factors are coupled in the presence of various environmental conditions, such as the co-existence of dust, high humidity, and high temperature in industrial environments.}.

\section{The Proposed Method of Flexible Resource Allocation}\label{sec:solution}

In this section, we focus on developing a solution to the problem defined in Section \ref{sec:framework}. For the purpose of initial investigation and for simplicity, we assume $D=1$ for the rate/throughput only. As a result, we can drop off the foot index of $d$, so that $a_{l}$ denotes the transmission rate of the $l$-th service,
and the ${\bf r}$ reduces to the scalar $r$ that is environment-dependent. Also, $\sum_{l=1}^L a_{l}$ is upper bounded by the system throughput as the function of the harsh environment. 

Firstly, we need to predict the system throughput, denoted by $\hat{r}$ based on the environmental conditions. To do this, we leverage deep learning owing to its superior capability to characterize the complicated underlying non-linear mapping function. Specifically, we denote by ${\bf e}$ the vector comprising of the environmental condition elements such as temperature, dust, humidity, electromagnetic interference, etc. Then we feed it as an input to a trainable model and the output is the prediction, i.e., $\hat{r}=\phi({\bf e})$ where $\phi(\cdot)$ is the to-be-learned underlying mapping function. Since we aim to predict the available amount of each resource, the learning problem becomes a multivariate regression problem. While several models can be employed, we employ a convolutional neural network (CNN) for the initial research purpose.

\begin{figure*}[!t]
\centering
\includegraphics[width=1\textwidth]{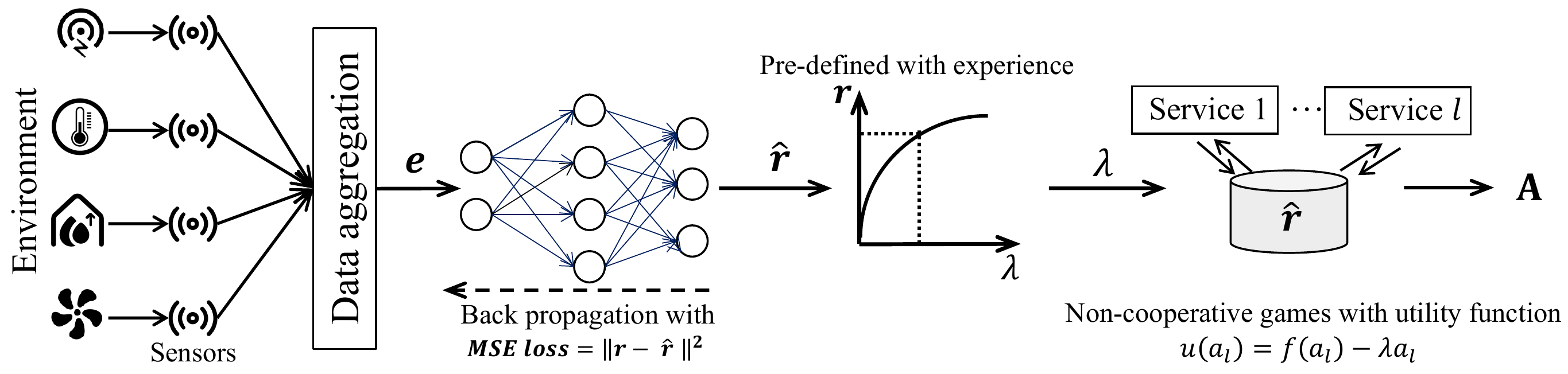}\vspace{-0.1in}
\caption{The proposed flexible resource allocation method.}
\label{strategy}
\end{figure*}

Next, we introduce a method to effectively maximize the system utility. Considering a centralized controller could be very expensive and fragile under extreme conditions, we assume the lack of cooperation and coordination among the services in the harsh environment. 
This naturally forms a non-cooperative scenario where each service needs to compete for the resource to optimize its own utility $f(a_l)$. In such a scenario, it would be natural to employ game theory as an ingredient. Recognizing this key, we model the resource allocation as a non-cooperative power control game where the services are the game players and their actions are characterized by resource allocation matrix $\bf{A}$. 

Note that in this game, if each service naively increases its transmission power $p_l$ for a higher rate $a_l$, more co-channel interference could be caused to other services. This in turn would suppress their rates and further induce them to continuously increase their own transmission power with diminishing or even no gains. On the other hand, if each service is very conservative in power consumption for interference control, it would also limit the rate. To characterize the trade-off between the two extremes, we introduce a pricing factor to the utility, which is regularized by a payment besides the transmission rate for each service, so that the new utility can be updated as $u(a_l) = w_l f(a_l) - \lambda p_l$, where $\lambda$ is the pricing factor and $a_l$ is the rate function of $p_l$. Moreover, the factor $\lambda$ needs to be carefully-designed. If $\lambda$ is small, each service tends to improve their utility by using higher power; and if $\lambda$ is large, the service would be more sensible to increasing power. In the harsh environment, since the throughput is affected by harsh conditions, dynamically adjusting the price factor according to the prediction $\hat{r}$ would be expected. 

\begin{algorithm}
\caption{\textit{Identify the Nash equilibrium of an established game}}
\label{alg} 
\begin{algorithmic}
\STATE set $j = 0$, $\epsilon > 0$ 
\STATE Randomly initialize the value of ${\bf A}(j)$ 
\REPEAT 
\STATE $j \leftarrow j+1$ 
\STATE Based on the game theory, each service competes its $a_l$ to optimize the utility function $u(a_l)$;\\
\STATE Calculate the difference value $ {\| {\bf A}(j) - {\bf A}(j-1) \|}_{\mathrm{F}} $
\UNTIL the value is less than $\epsilon$.   \\
\RETURN ${\bf A}(j)$
\end{algorithmic}
\end{algorithm}

By using the game-theory model, the system can achieve the Nash equilibrium from which each rational service is unwilling to deviate. The method to achieve the Nash equilibrium is provided in Algorithm \ref{alg} for the resource allocation game. In addition, we also provide an overview of the solution in Fig.~\ref{strategy}. The environmental conditions are sensed by the distributed sensors, then aggregated as the input of the environment-resource prediction model. The prediction model outputs the available network resources for scheduling. Note that the available resources are adjusted dynamically to address the environmental dynamic changes in the harsh environment. Then the resources are flexibly allocated to each service based on a game-theory method with the pre-defined price factor. To this end, we achieve the environment-aware and self-optimization of the services.

\section{A Case Study: An Experiment and the Result}\label{sec:example}

To show the effectiveness of the proposed framework introduced in Section \ref{sec:framework}, we conduct a case study by applying the proposed solution to a wireless network in presence of a power station under harsh conditions.

\subsection{The Considered Scenario}

The wireless network in the area around large-scale power station usually suffers from the higher noise level and lower transmission rate due to the presence of the high electromagnetic interference (EMI). In particular, the high-power electronic devices, voltage load and switch operations could significantly change the nearby field strength, causing high noise and interference to the receiver, and thus further limiting the system throughput. Thus, the wireless network that still provides coverage and service applications in this scenario would need careful design so as to meet the QoS requirements. 

In the experiment, we collected a dataset of measurements from China Unicom's 5G cellular network close to a power station in the city of Qingdao in Shandong Province of China, where the strong EMI widely affects network performance. The dataset includes 416 data points, each with a total of 32 continuous environmental factors. These 32 factors consist of 30 EMI strength indicators denoted by the radiation power density on 30 uniformly distributed positions, the environment temperature and the humidity. At each position, we used Rohde $\&$ Schwarz ESRP to measure the EMI strength on the 3.5 GHz -- 3.6 GHz frequency band. To capture abundant types of harsh environments, the measurements were collected every eight hours, i.e., three times per day, for a total of non-consecutive 140 days from January to October, 2019. 
Finally, in the lab, we labeled each data point with the corresponding throughput that was also collected from the network.

\subsection{Deep Learning for Available Resource Prediction}

\begin{figure*}[t]
\centering
\includegraphics[width=0.8\textwidth]{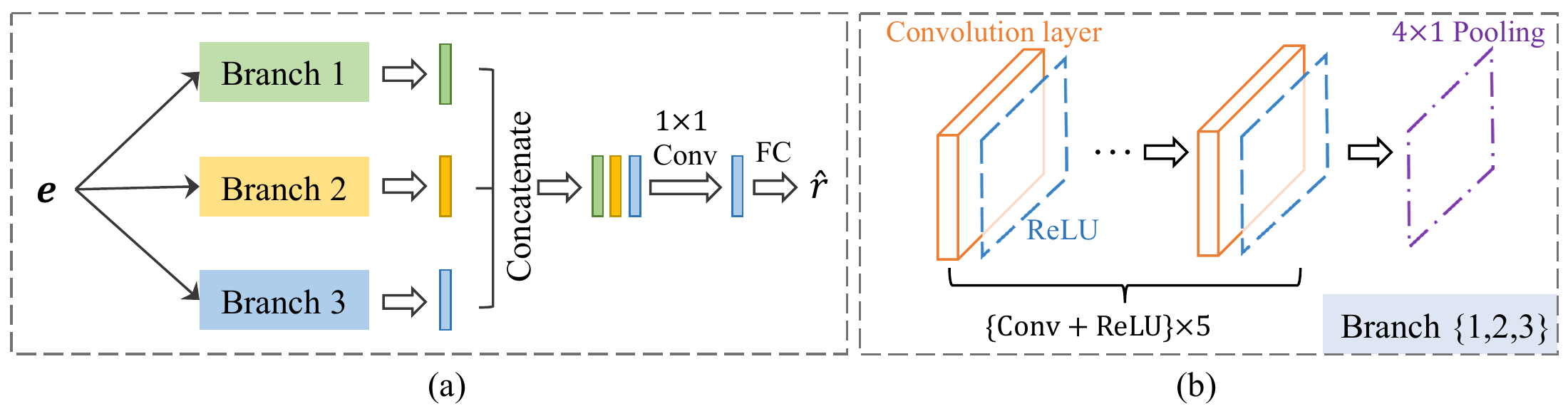}\vspace{-0.1in}
\caption{The deep neural network for resource prediction. Conv: convolutional layer; FC: fully connected layer. (a) The structure of the proposed network. (b) The structure of each of the three branches.}
\label{fig:dl_model}
\end{figure*}

As introduced in Section \ref{sec:solution}, we employ deep learning to predict the maximum throughput for multiple environment factors. In particular, we implement a 1-D CNN-based three-branch deep neural network, as shown in Fig.~\ref{fig:dl_model} (a), as we expect to capture the spatial correlation over the space. 
To improve the ability of extracting multi-scale information, similar to \cite{marnerides2018expandnet}, we build a three-branch CNN-based neural network. The three branches are deployed with different kernel size to enrich the diversity of receptive multi-scale features.
We demonstrate the structure of one branch in Fig.~\ref{fig:dl_model} (b). Each branch consists of 5 convolutional layers of the same size, each followed by a ReLU function as non-linear activation. The number of kernels for each convolutional layer is 64, 64, 16, 16 and 4, respectively. A 4$\times$1 max-pooling layer is added at the end of the branch to down-sample the data stream. Note that three branches are deployed with different kernel size to enrich the diversity of receptive multi-scale features. Specifically, $1\times 1$, $3\times 1$, and $5\times 1$ 1-D kernels, each with stride 1, and the padding sizes 0, 1, 2 are considered for the three parallel branches, respectively. As a result, the dimension of the signal right after the pooling is given by $8\times 1 \times 4$ for each branch. Next, we concatenate them along the last dimension,  and then use 3 kernels with the size of $1\times 1$ in another convolutional layer, and finally followed by a $24\times 1$ fully connected layer to produce the prediction. Considering this is a regression problem, we employ the mean squared error as the loss function.

To train the model defined above, we choose the model hyper-parameters as: the epoch number = 50, the batch size = 8, the Adam optimizer \cite{kingma2014adam} with $\beta_1 = 0.9$ and $\beta_2 = 0.999$, the learning rate = $4 \times 10^{-4}$, and we use the Xavier initialization \cite{glorot2010understanding} method to initialize the weights. Once the model is established, we split the dataset to non-overlapping training and testing subsets according to $75\%/25\%$ splitting criteria, and thus 312 data points are used for training and the other 104 data points are used for testing. 



\begin{figure*}[!t]
\begin{minipage}[t]{0.5\textwidth}
\centering
{\includegraphics[width=1\textwidth]{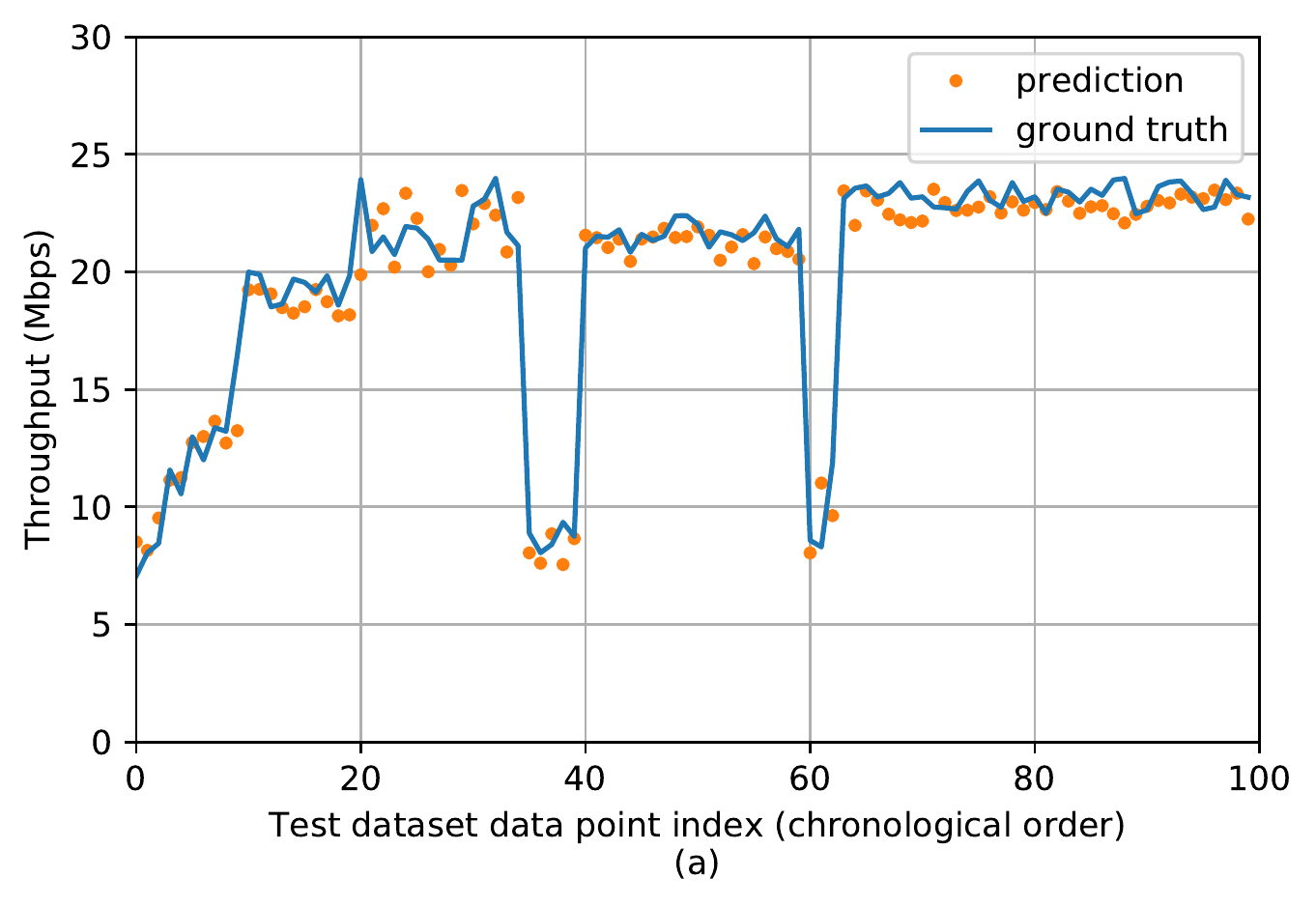}}
\label{fig:env-pre}
\end{minipage}
\begin{minipage}[t]{0.5\textwidth}
\centering
{\includegraphics[width=1\textwidth]{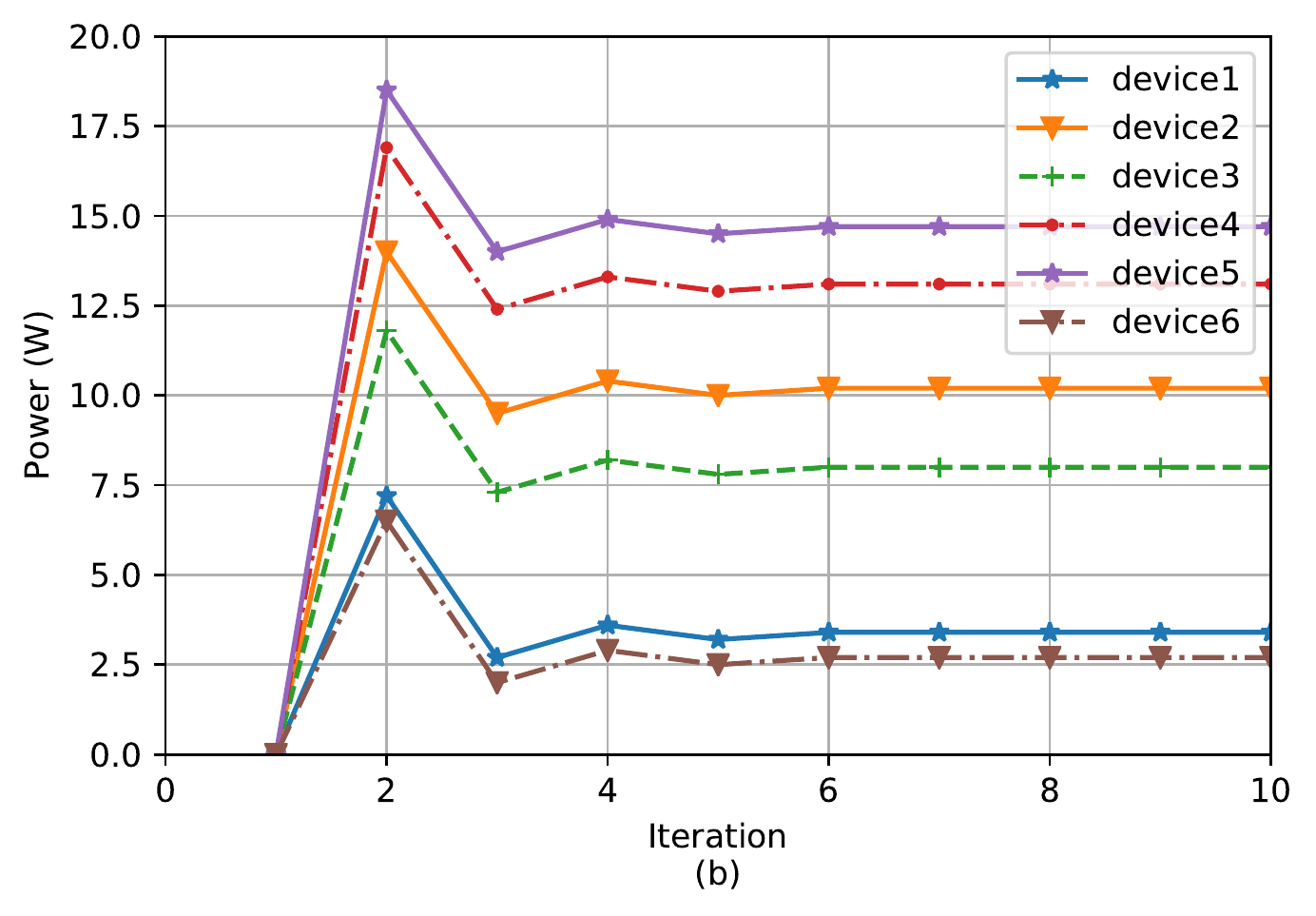}}
\label{iter}
\end{minipage}\vspace{-0.2in}
\begin{minipage}[t]{0.5\textwidth}
\centering
{\includegraphics[width=1\textwidth]{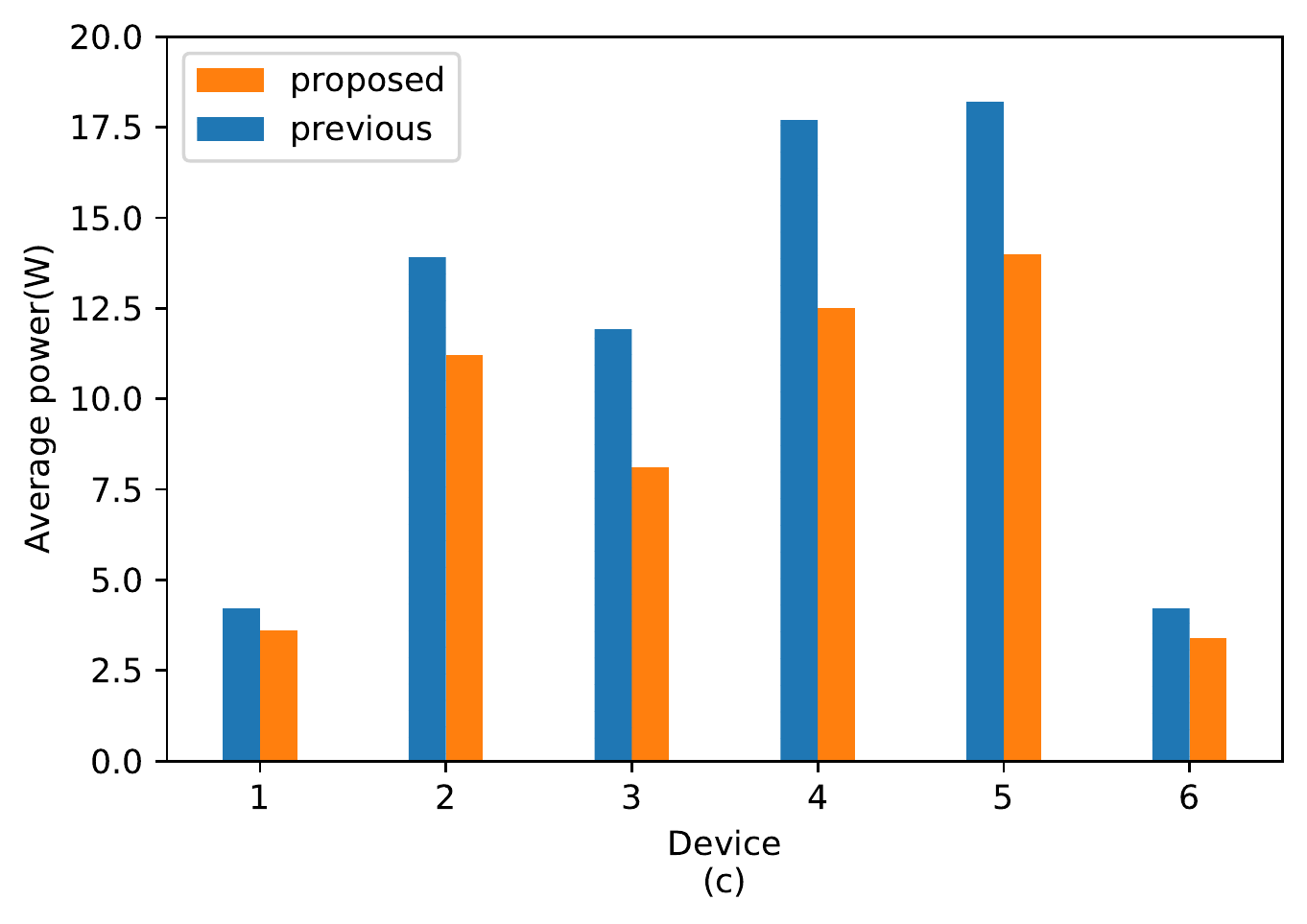}}
\label{power}
\end{minipage}
\begin{minipage}[t]{0.5\textwidth}
\centering
{\includegraphics[width=1\textwidth]{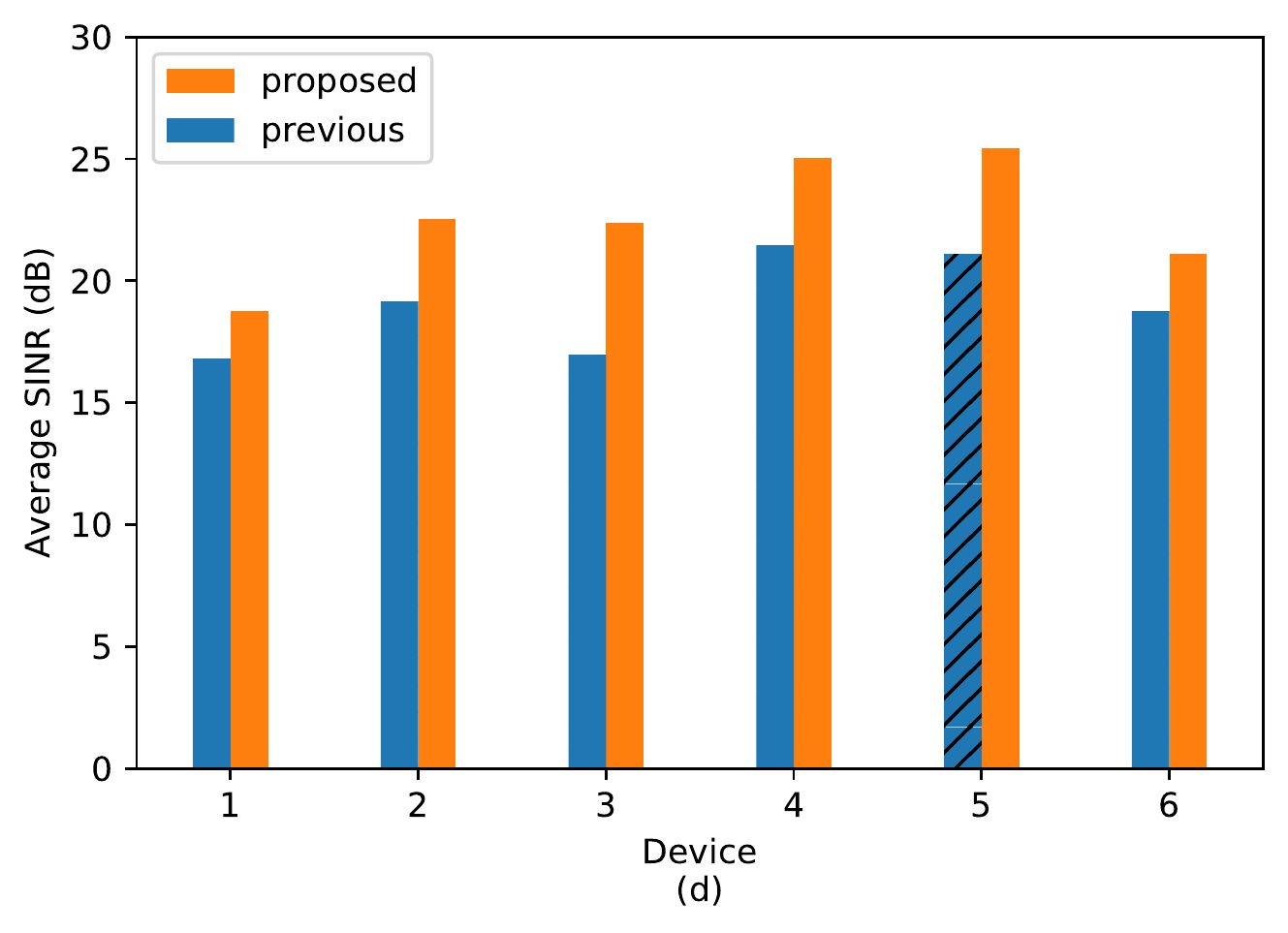}}
\label{sinr}
\end{minipage}\vspace{-0.2in}
\caption{(a) The predicted throughput performance of the proposed environment-resource prediction model on the test dataset (sorted in an ascending order w.r.t. their ground truth).
(b) The utility of each device over the iterations of the algorithm.
(c) The average-SINR performance comparison between the proposed method and the method in \cite{nathani2012policy}. (d) The average-power performance comparison between the proposed method and the method in \cite{nathani2012policy}.}
\label{fig4}
\end{figure*}

To evaluate the performance of the model defined above, we show the predicted results of the test dataset in Fig.~\ref{fig4} (a). We give the data samples in chronological order. It can be seen that the predictions well match with the ground truth in general. When rapid changes occur (data point index around 40 and 60), the model well predicts the thoughputs. Indeed, we achieve 0.955 for the R-Squared, 0.943 for the Root Mean Squared Error (RMSE), and $3.5\%$ for the Relative Error. 

\subsection{Evaluation of the Proposed Resource allocation Method}

After obtaining $\hat{r}$, we substitute $r_d$ in the formulated problem with $\hat{r}$.  
Then, our goal is to maximize the function defined as $u(a_l)$ in replace of $f(a_l)$ by exploring the optimal network resource allocation. For the purpose of proof of concept, we consider a simple setting of $f(a_l)=a_l=B\log(1+\textrm{SINR})$ where $B$ is the bandwidth and SINR denotes the signal-to-interference-and-noise ratio. Also, we assume the SINR only depends on the second-order statistics of the channels and the associated transmit power\footnote{This assumption makes the problem easily tractable and can be used as a baseline for the proof of concept. In practice, it can also be realized by using the massive-antenna arrays owing to channel hardening. In terms of the required techniques to optimize the utility, admittedly, this assumption would allow a wider range of optimization tools in addition to the game-theory-based method that we proposed in this paper.}. For the evaluation purpose, we also use SINR as the metric to evaluate the QoS. 
In addition to $f(a_l)$, the utility function of each device also considers its QoS, captured by the variable $p_l$. 


For the purpose of initial investigation, we assume 6 services and co-channel interference existing in the network. To evaluate the performance of the proposed dynamic resource allocation method, we compare the proposed method to the one investigated in \cite{nathani2012policy} that used a static resource allocation method, ignored the dynamic environment constraints, the power of the device was manually adjusted and stayed constantly over a unmanned period. To combat with the environment EMI interference, \cite{nathani2012policy} set the power to a relatively high level and the network resource was allocated if available and the request is rejected otherwise. In contrast, our proposed method dynamically adjusts the power of the devices according to the predicted throughput in presence of dynamic environment conditions. That is, based on the predicted throughput constraint $\hat{r}$, we tune the factor $\lambda$ and employ Algorithm 1 proposed in Section \ref{sec:framework}.B. 

In Fig.~\ref{fig4} (b), we evaluate the power level of the 6 devices over the algorithm iterations to achieve the Nash equilibrium. It can be seen that the algorithm tends to converge after the $6$-th iteration. 
In fact, it is well known that the Nash-equilibrium point based on the pure strategy always exists \cite{han2012game}. Thus, if the number of $L$ is large, it might take longer time for the algorithm to converge.

Finally, we evaluate the average SINR and average power. In particular, we 
calculate the average rate and the average SINR of the 104 data points in the test dataset. In Fig.~\ref{fig4} (c), it can be seen that the proposed scheme requires lower average power than the previous method by 22.6\% on average; in Fig.~\ref{fig4} (d), the proposed algorithm achieves higher average SINR than the previous algorithm by 18.5\% on average. 



\section{Future Work Directions}

As the initial investigation of the resource allocation under harsh conditions, we simplify the problem with a few assumptions that might not, if not impossible, fully characterize the environment dynamics. In this section, we suggest three components that we believe are worthy of further investigation.

\emph{Channel access management:} The ultra density of IoT devices in harsh environments, such as massive sensors and communication nodes in industrial production environments, brings challenges related to network congestion, networking and storage architecture, and efficient data communication protocols under limited resources. These access mechanisms should be cost-effective and balance between the QoS requirements and the limited network resources. Also, due to the environment dynamics, the access management architecture should dynamically adapt to the resource status as well to meet the QoS requirements.



\emph{Effective energy management:} The harsh environment generally reduces the battery lifespan, and thus the energy efficiency turns from an optional factor to a necessary design factor, so as to provide reliable support. Ideally, the energy management method should consider power allocation, energy harvesting and scheduling in dynamic networks \cite{liu2019intelligent}. In some remote environments, the devices have to be powered by the battery due to the lack of power supply equipment. Hence, novel energy harvesting techniques would be necessary to extend the lifespan during the long unmanned period.

\emph{Integration of sensing and communication capability:} Due to the network requirement such as high reliability, self-healing, and autonomy in harsh conditions, it is essential to enhance communication and networking capabilities by obtaining ambient environmental awareness. One popular idea is to integrate wireless sensing capability into wireless communication \cite{9606831}, so that the environmental awareness could be extracted from the signal reflected from the surrounding scatters. Currently, the coordination between resultant sensing and communication services is still widely under investigation, particularly on the scheduling and resource allocation strategies of sensing and communications to maximize the overall network performance.

\section{Conclusion}

The harsh environment brought considerable challenges to the network resource management. In this article, considering two main challenges, i.e., the uncertainty of the environmental impact and the service requirement of a flexible adaptation mechanism, we proposed a flexible resource management architecture that extended the functionalities of the physical plane, the resource plane, and the application plane from traditional architectures. Then, to incorporate the complicated environmental impact on network resources as an ingredient, we developed a deep-learning-based environment-resource prediction model to predict the resource constraints. With the predicted resource conditions, we finally scheduled the network resources via game theory to maximize the system utility. We also presented a case study to illustrate the effectiveness of the proposed architecture. Finally, we suggested three promising directions of the network resource management, which are worthy of further investigation in the future.


\newpage

\section{Biographies}

\vspace{11pt}

\begin{IEEEbiographynophoto}{Jiaqi Zou}
is currently pursuing her Ph.D degree at Beijing University of Posts and Telecommunications. She received her B.S. degree from Beijing University of Posts and Telecommunications in 2020.
Her current research interests include wireless communication and computer vision.
\end{IEEEbiographynophoto}

\begin{IEEEbiographynophoto}{Rui Liu}
is with San Jose State University, California as an Associate Professor at the Department of Economics. She received her Ph.D. in Economics from the University of California, Irvine, US, in 2013. Her research interests include econometrics, Bayesian statistics, time series, and applied machine learning.
\end{IEEEbiographynophoto}

\begin{IEEEbiographynophoto}{Chenwei Wang}
is with Google LLC, Mountain View, California as a Data Scientist, Engineering. He received his Ph.D. in Electrical and Computer Engineering from the University of California, Irvine, US, in 2012, and had been with DOCOMO Innovations Inc., Palo Alto, California as a Research Engineer from 2013 to 2021.
\end{IEEEbiographynophoto}

\begin{IEEEbiographynophoto}{Yuanhao Cui}
received Ph.D. degree at Beijing University of Posts and Telecommunications. He is the CTO and co-founder of two startup companies. He organized a number of ISAC workshops. He is the Secretary of IEEE ComSoc ISAC-ETI and CCF Science Communication Working Committee. His research interests is ISAC.
\end{IEEEbiographynophoto}

\begin{IEEEbiographynophoto}{Zixuan Zou}
is currently pursuing his M.S. degree at Beijing University of Posts and Telecommunications. He received his B.S. degree from Beijing University of Posts and Telecommunications in 2021.
His current research interests include wireless communication and computer vision.
\end{IEEEbiographynophoto}

\begin{IEEEbiographynophoto}{Songlin Sun}
is currently a professor at Beijing University of Posts and Telecommunications. He received his Ph.D. degree from Beijing University of Posts and Telecommunications in 2003. His research interests include video codec, signal processing, wireless communication, and deep learning.
\end{IEEEbiographynophoto}

\begin{IEEEbiographynophoto}{Koichi Adachi} 
received his Ph.D. degree in engineering from Keio University, Japan, in 2009, and he is currently an associate professor at the University of Electro-Communications, Japan. His research interests include machine learning, the Internet-of-Things, wireless sensor network, and radio resource management. He is a senior member of IEEE.
\end{IEEEbiographynophoto}

\vfill

\end{document}